\pdfoutput=1

\documentclass{article}

\usepackage{microtype}
\usepackage{graphicx}
\usepackage{subfigure}
\usepackage{booktabs}
\usepackage{amsmath,amssymb,amsthm}
\usepackage{algorithm}
\usepackage{algorithmic}
\usepackage{multirow}
\usepackage{hyperref}
\usepackage{orcidlink}
\usepackage{placeins}
\usepackage{caption}

\usepackage[accepted]{mlsys2025}

\mlsystitlerunning{CodeCRDT: Observation-Driven Coordination for Multi-Agent LLM Code Generation}

\begin{document}

\twocolumn[
\mlsystitle{CodeCRDT: Observation-Driven Coordination for Multi-Agent LLM Code Generation}


\mlsyssetsymbol{equal}{*}

\begin{mlsysauthorlist}
\mlsysauthor{Sergey Pugachev (\texttt{sergey@spugachev.com}) \orcidlink{0009-0008-5134-6411}}{}
\end{mlsysauthorlist}

\mlsyskeywords{Conflict-Free Replicated Data Types, Multi-Agent Systems, Large Language Models, Distributed Synchronization, Strong Eventual Consistency, Parallel Code Generation}

\vskip 0.3in

\begin{abstract}
Multi-agent LLM systems fail to realize parallel speedups due to costly coordination. We present CodeCRDT, demonstrating an \textbf{observation-driven coordination pattern}: agents coordinate by monitoring a shared state with observable updates and deterministic convergence, rather than through explicit message passing. Agents observe edits, skip completed work, integrate context, and avoid conflicts. We instantiate this pattern using Conflict-Free Replicated Data Types (CRDTs), which provide strong eventual consistency (SEC) enabling lock-free, conflict-free concurrent code generation.

Evaluation across 600 trials (6 tasks, 50 runs per mode) using Claude Sonnet 4.5 reveals both benefits and trade-offs: parallel coordination achieves up to 21.1\% speedup on some tasks but incurs up to 39.4\% slowdown on others (p$<$0.001). Observed slowdowns result from confounded factors---code generation volume inflation (82--189\% for complex tasks), LLM latency variability, and observation/coordination overhead---that current measurements cannot isolate. Strong eventual consistency guarantees 100\% convergence with zero merge failures (no manual conflict resolution needed), though preliminary inspection suggests 5--10\% semantic conflicts. Parallel agents optimize runtime performance (+25\%, d=1.51) but degrade code quality (-7.7\%, d=-0.71) as assessed via LLM-based evaluation (limitation: no human validation baseline). Performance depends on task characteristics including component interdependencies.
\end{abstract}
]


\section{Introduction}
\label{sec:introduction}

Multi-agent LLM systems promise parallel speedups but existing approaches fail: sequential frameworks (ChatDev~\cite{chen2024chatdev}, MetaGPT~\cite{hong2023metagpt}) employ waterfall/pipeline workflows precluding concurrent execution; lock-based coordination introduces O(N$\times$L) contention (N agents competing for L locks); git-style branching defers 15--30\% conflicts to costly merge-time resolution. The core challenge is enabling \textbf{lock-free concurrent editing while maintaining strong consistency guarantees}---a problem extensively studied in distributed systems but not previously applied to autonomous multi-agent LLMs.

We apply \textbf{observation-driven coordination}---a decades-old pattern from distributed systems (Linda tuplespaces~\cite{gelernter1985linda}, blackboard architectures~\cite{erman1980hearsay}, stigmergy~\cite{theraulaz1999stigmergy})---to autonomous LLM agents. In this pattern, agents coordinate through monitoring shared state rather than explicit message passing, observing edits, skipping completed work, and avoiding conflicts without centralized task assignment. The pattern requires a shared state substrate with three properties: (1) \textbf{observable updates} (agents can subscribe to state changes), (2) \textbf{deterministic convergence} (all agents eventually observe consistent state), and (3) \textbf{monotonic progress} (no rollbacks invalidating completed work). These properties can be provided by diverse consistency models---strong eventual consistency (SEC) via CRDTs~\cite{shapiro2011crdts}, centrally arbitrated sequential consistency via OT (Operational Transform) systems, or strong consistency via replicated logs---each with different cost/guarantee trade-offs. We instantiate using \textbf{CRDTs} for SEC, enabling lock-free deterministic convergence with zero character-level merge conflicts. While prior work applied observation-driven coordination to deterministic processes or humans, we demonstrate its viability for \textbf{stochastic LLM agents} and characterize agent-specific failure modes (semantic conflicts, task-dependent performance variability) absent in traditional contexts.

\textbf{Empirical Findings}: Evaluation across 600 trials (6 tasks, 50 runs per mode, Claude Sonnet 4.5) reveals variable outcomes: raw response times show 21.1\% speedup to 39.4\% slowdown. \textbf{Normalized time analysis controlling for code volume reveals parallel is faster per character for 5/6 tasks} (11--52\% speedup), demonstrating that apparent slowdowns result from code generation volume (parallel produces 82--189\% more code with optimizations) rather than coordination inefficiency. Only highly coupled tasks show true coordination overhead. Strong eventual consistency guarantees 100\% convergence but preliminary inspection suggests 5--10\% semantic conflicts (duplicate declarations, type mismatches) require post-generation reconciliation. Parallel agents optimize runtime performance (+25\%, d=1.51) correlating with code volume, but degrade code quality (-7.7\%, d=-0.71). Results validate observation-driven coordination effectiveness while revealing emergent behavior: parallel agents optimize locally, producing more robust but verbose code.

\subsection{Contributions}

\begin{enumerate}
\item Formalization and implementation of observation-driven coordination for stochastic LLM agents with provable safety (TODO-claim protocol ensuring at-most-one agent per task under strong eventual consistency)
\item Empirical evaluation across 600 trials characterizing performance trade-offs: parallel coordination shows task-dependent outcomes, with task structure influencing scalability---a finding not characterized in prior shared-state coordination work
\item Demonstration that LLM agents exhibit distinct failure modes (5--10\% semantic conflicts, quality-performance trade-offs) requiring reconciliation despite character-level convergence
\item Deployment heuristics and identification of confounded factors (code generation volume, LLM latency variability, O(N$\times$U) observation overhead with N agents monitoring U updates) influencing when parallel coordination succeeds vs. fails
\end{enumerate}

This work applies decades of shared-state coordination research to stochastic LLM agents, characterizing both successes and limitations of parallel coordination patterns.

\section{Background and Related Work}
\label{sec:background}

\subsection{Shared-State Coordination}

Linda tuplespaces~\cite{gelernter1985linda} pioneered coordination via shared associative memory; blackboard architectures (Hearsay-II~\cite{erman1980hearsay}) coordinate agents via shared problem-solving state. Stigmergic coordination~\cite{theraulaz1999stigmergy} in multi-robot systems uses environment modifications (virtual pheromones) for indirect agent coordination. These approaches coordinate via observation but lack: (1) formal guarantees for deterministic convergence under concurrent writes (Linda uses locking, blackboards use centralized serialization, stigmergy is best-effort), (2) application to LLM agents with stochastic behavior and semantic reasoning, (3) empirical characterization of performance across diverse task structures. \textbf{Gap}: Prior work has not formalized observation-driven coordination for LLM agents with provable convergence guarantees and empirical scalability analysis across varied tasks.

\subsection{CRDTs and Eventual Consistency}

Shapiro et al.~\cite{shapiro2011crdts} formalized strong eventual consistency via commutative operations. Modern implementations (Yjs~\cite{yjs}, Eg-walker~\cite{gentle2025egwalker}) enable human collaborative editing at scale. Production IDEs use centralized OT requiring server arbitration~\cite{vscode2020liveshare}. \textbf{Gap}: Prior CRDT work targets human-human collaboration; applying SEC to autonomous LLM agent coordination requires new protocols and characterization of agent-specific failure modes.

\subsection{Multi-Agent LLM Systems}

Sequential frameworks (ChatDev~\cite{chen2024chatdev}, MetaGPT~\cite{hong2023metagpt}) employ waterfall/pipeline execution precluding concurrent speedups; orchestrator-based systems introduce centralized bottlenecks with explicit task assignment; decentralized approaches lack formal consistency guarantees. \textbf{Gap}: To our knowledge, no prior system combines observation-driven coordination with strong eventual consistency for concurrent LLM-based editing.

\subsection{LLM Code Generation}

LLM-based code generation shows significant productivity improvements (e.g., GitHub Copilot improves productivity 55.8\%~\cite{peng2023copilot}); multi-agent approaches show promise through role distribution. \textbf{Gap}: Prior work has not empirically characterized when parallel LLM agents succeed vs. fail based on task structure.

\subsection{Positioning}

CodeCRDT builds on decades of shared-state coordination research (Linda, blackboard systems, stigmergy) but makes three novel contributions for LLM agents: (1) formal TODO-claim protocol with provable at-most-one-winner safety under strong eventual consistency, (2) empirical characterization revealing how task structure influences parallel coordination effectiveness---a finding not characterized in prior work, (3) demonstration that LLM agents' stochastic behavior and semantic reasoning introduce failure modes (semantic conflicts, quality-performance trade-offs) distinct from traditional coordination contexts.

\begin{table*}[t]
\caption{CodeCRDT vs. prior shared-state coordination approaches}
\label{tab:comparison}
\vskip 0.15in
\centering
\small
\begin{tabular}{lllll}
\toprule
\textbf{Property} & \textbf{Linda/Tuplespaces} & \textbf{Blackboard} & \textbf{Stigmergy} & \textbf{CodeCRDT} \\
\midrule
Coordination & Shared tuple space & Shared blackboard & Environment & Shared CRDT \\
Agent Type & Deterministic & Knowledge sources & Reactive robots & \textbf{Stochastic LLMs} \\
Consistency & Atomic/locks & Sequential & Best-effort & \textbf{Strong Eventual} \\
Observable Updates & Polling/blocking & Event notifications & Sensing & \textbf{CRDT observe()} \\
Convergence & N/A (atomic) & N/A (centralized) & Emergent & \textbf{Deterministic} \\
Conflict Resolution & Locking/queuing & Sequential KS & Emergent & \textbf{Automatic} \\
Safety Guarantees & Via locking & Via serialization & None & \textbf{Formal (SEC)} \\
Scalability Analysis & Theoretical & Case-specific & Simulation & \textbf{Empirical} \\
Document Editing & No & No & No & \textbf{Yes} \\
Semantic Conflicts & N/A & KS logic & N/A & \textbf{5--10\% measured} \\
\bottomrule
\end{tabular}
\end{table*}

\section{Method: Observation-Driven Coordination}
\label{sec:method}

\subsection{Design Principles}

CodeCRDT embodies: (1) \textbf{Coordination via State Observation}---agents coordinate through monitoring shared state with observable updates and deterministic convergence; (2) \textbf{Specialization}---role-based agents with formal protocols; (3) \textbf{Deterministic Convergence}---underlying substrate guarantees all agents eventually observe consistent state; (4) \textbf{Extensibility}---pattern generalizes to substrates providing required properties.

\subsection{Substrate Requirements}

Observation-driven coordination requires three properties from the shared state substrate:
\begin{itemize}
\item \textbf{Observable updates}: agents can subscribe to state changes via event notifications
\item \textbf{Deterministic convergence}: all agents eventually observe consistent state without manual conflict resolution
\item \textbf{Monotonic progress}: no rollbacks invalidating completed agent work
\end{itemize}

Different consistency models provide these properties with varying guarantees and costs:

\begin{itemize}
\item \textbf{Strong Eventual Consistency (SEC) via CRDTs}: Decentralized, partition-tolerant, lock-free; automatic conflict resolution via commutative operations; eventual convergence with bounded staleness; O(N$\times$O) metadata overhead for N agents, O operations
\item \textbf{Sequential Consistency via Operational Transformation (OT)}: Centrally arbitrated, lower latency for small N; requires server availability; single point of failure; simpler conflict resolution than CRDTs
\item \textbf{Strong Consistency via Replicated Logs}: Linearizable reads, immediate consistency; higher coordination cost (Raft/Paxos); predictable latency; requires quorum for progress
\end{itemize}

\textbf{Implementation Choice}: We instantiate using Yjs CRDTs via centralized Hocuspocus WebSocket relay (SEC with simplified deployment). This provides lock-free coordination with partition tolerance while centralizing persistence and observability. Our evaluation isolates coordination pattern characteristics (observation overhead, task-dependent performance) from CRDT-specific trade-offs; alternative substrates would exhibit different metadata overhead and convergence latency but similar task-dependent behavior patterns.

\subsection{System Architecture}

\begin{figure*}[!ht]
\centering
\includegraphics[width=0.8\textwidth,height=0.38\textheight,keepaspectratio]{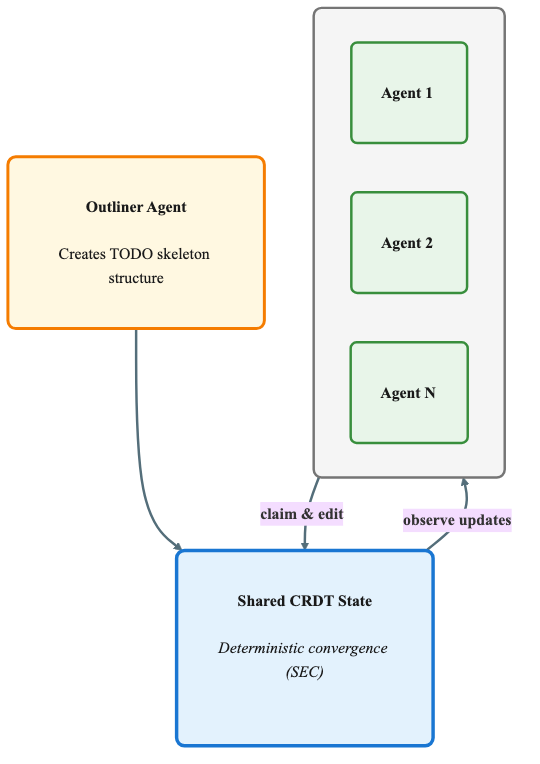}
\caption{CodeCRDT System Architecture. Observation-driven coordination pattern showing lock-free parallel agent execution. The Outliner Agent creates TODO skeleton in Shared CRDT State; Implementation Agents coordinate by observing updates and claiming TODOs via optimistic write-verify protocol. Strong eventual consistency (SEC) guarantees deterministic convergence with zero character-level conflicts.}
\label{fig:architecture}
\end{figure*}

Core components include: (1) \textbf{Inference Service}---backend managing task queue and agent lifecycle; (2) \textbf{Yjs Document}---shared CRDT state accessible via WebSocket; (3) \textbf{Agents}---LLM-powered workers (outliner, implementation) with CRDT write access; (4) \textbf{TODO Observer}---real-time scanner detecting TODOs as insertion points; (5) \textbf{Frontend}---React/Monaco editor for live visualization. Figure~\ref{fig:architecture} shows the complete system architecture.

Agents communicate exclusively through three shared CRDT types (Table~\ref{tab:crdttypes}).

\begin{table}[h!]
\caption{Yjs CRDT types and coordination invariants}
\label{tab:crdttypes}
\vskip 0.15in
\centering
\small
\begin{tabular}{ll}
\toprule
\textbf{Yjs Type} & \textbf{Purpose and Key Invariants} \\
\midrule
\textbf{Y.Text} & Code document; character-level convergence \\
& Deterministic ordering via operation ID \\
\midrule
\textbf{Y.Map} & TODO assignments \& agent coordination \\
& At-most-one assignedTo per TODO \\
& LWW semantics per key \\
\midrule
\textbf{Y.Array} & Append-only audit trail \\
& Causally ordered message history \\
\bottomrule
\end{tabular}
\end{table}

\FloatBarrier

\subsection{Agent Roles}

\textbf{Outliner} generates TypeScript/React skeletons with TODO placeholders. \textbf{Implementation} agents scan for unassigned TODOs, claim via CRDT protocol, and fill with working code. \textbf{Evaluator} analyzes final code for quality, architecture, performance, and accessibility (0--100 score).

\subsection{TODO Claim Protocol}

Agents claim work via optimistic write-verify on shared Y.Map (LWW register semantics per key):

\begin{enumerate}
\item \textbf{Scan}: Read Y.Map, filter \texttt{\{status='pending', assignedTo=null\}}
\item \textbf{Claim}: Write \texttt{TODO\_k.assignedTo $\leftarrow$ agentId}
\item \textbf{Verify}: After CRDT sync delay (50~ms), re-read \texttt{TODO\_k.assignedTo}
\item \textbf{Proceed}: If \texttt{assignedTo == agentId}, claim succeeded; else, retry
\end{enumerate}

\textbf{Safety Invariant}: Yjs Y.Map uses LWW register semantics---concurrent writes resolve via (logical clock, clientID) lexicographic ordering. All replicas converge to same winner deterministically.

\clearpage

\begin{figure*}[!ht]
\centering
\includegraphics[width=0.7\textwidth]{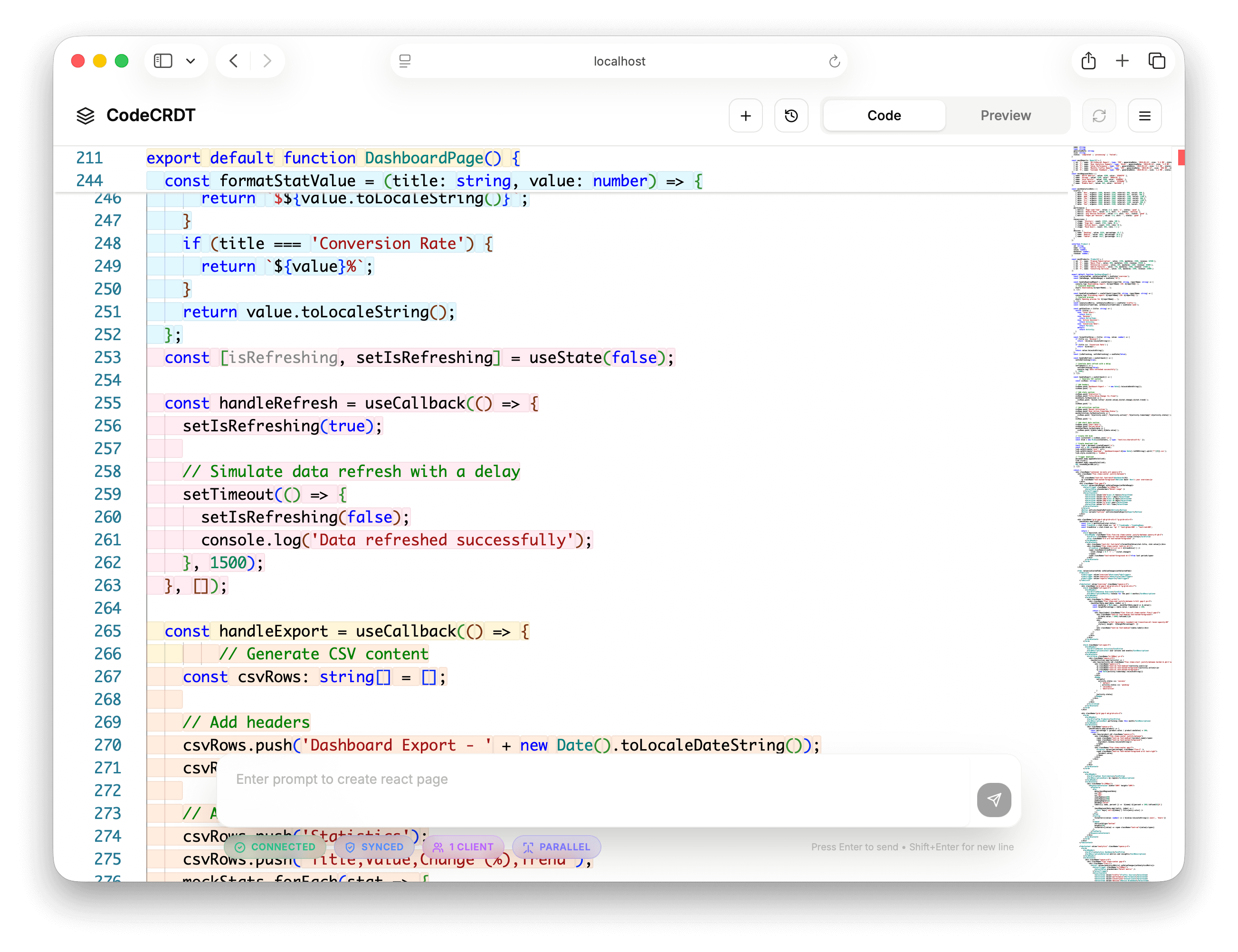}
\caption{Monaco Editor showing concurrent multi-agent code generation with colored cursors indicating different agents editing TODOs while CRDT synchronization maintains a single converged document.}
\label{fig:editor}
\end{figure*}

\begin{figure*}[!ht]
\centering
\includegraphics[width=0.7\textwidth]{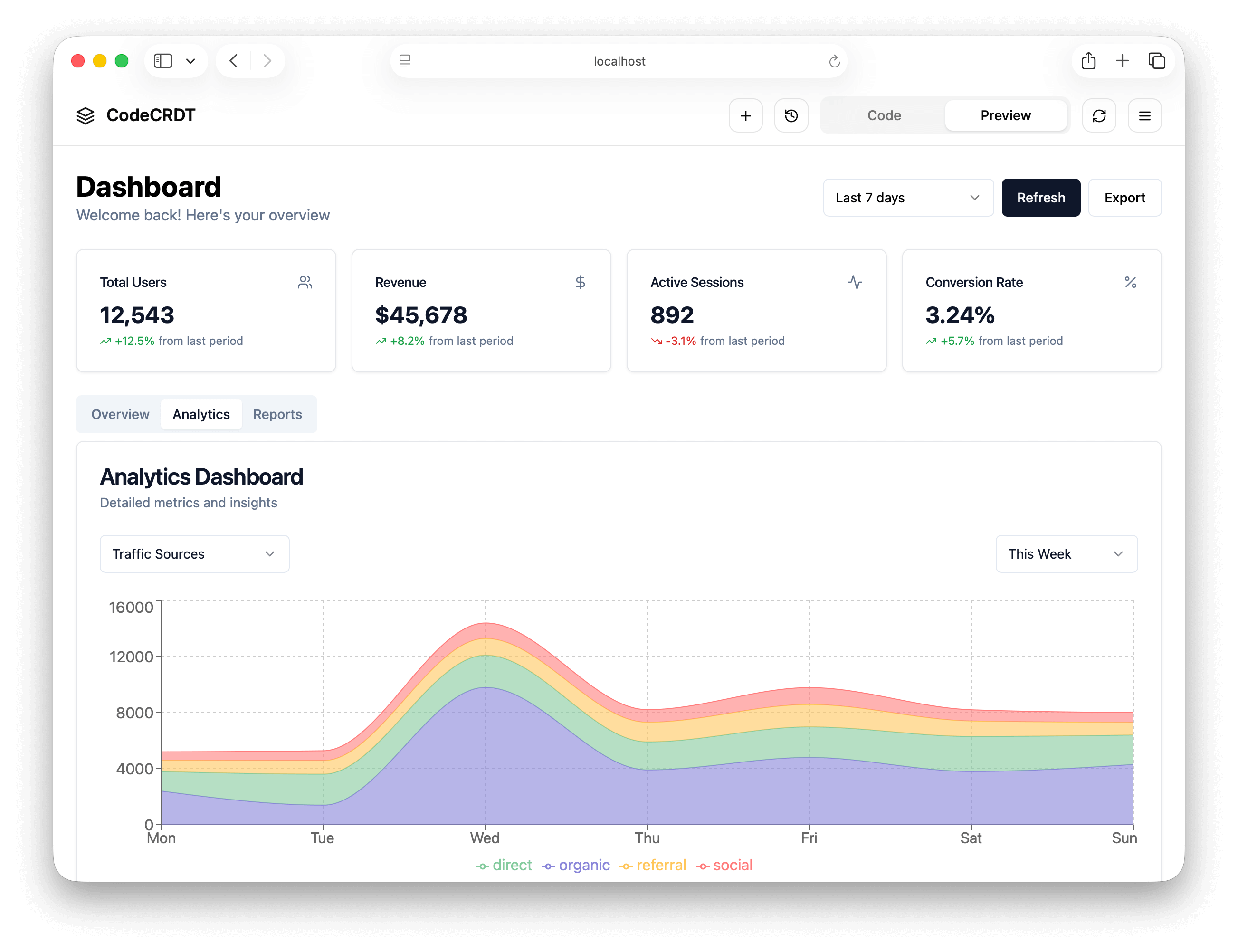}
\caption{Live preview rendering the evolving React application in real time, demonstrating end-to-end execution of agents' collaborative edits.}
\label{fig:preview}
\end{figure*}

\clearpage

\section{Implementation: Agent Coordination}
\label{sec:implementation}

\subsection{Coordination Protocol}

Implementation agents start concurrently after outliner; claim TODOs via CRDT map operation (\texttt{TODO.assignedTo = agentId}); evaluator waits for completion count. No explicit message passing---CRDT state serves as coordination substrate.

\subsection{Observation-Driven Adaptation}

Agents subscribe to CRDT events for real-time state updates, employing: (1) \textbf{Completed Work Detection}---skip implemented TODOs, (2) \textbf{Context Integration}---incorporate new imports/types, (3) \textbf{Naming Alignment}---match code style, (4) \textbf{Conflict Avoidance}---back off when editing regions overlap. CRDT updates propagate with median 50~ms latency (95th percentile: 200~ms).

\textbf{Scalability}: Observation-adaptation introduces O(N$\times$U) overhead (N agents, U update frequency). Each CRDT update triggers callbacks in all N agents. Observed slowdowns result from confounded factors: observation/coordination overhead, code generation volume inflation (82--189\% for complex tasks), and LLM latency variability ($\pm$21.57~s). Tasks achieving speedup (up to 21.1\%) minimize these overheads through isolated TODO execution.

\subsection{Conflict Resolution}

\textbf{Character-Level (Automatic)}: Yjs deterministically merges concurrent edits via operation ID total order (e.g., two agents inserting at same line $\rightarrow$ both preserved in deterministic order; edits interleave but never require manual conflict resolution). \textbf{Guarantee}: Strong eventual consistency---all agents converge to identical state with zero merge failures (0\% character-level conflicts means no manual conflict resolution needed, not that edits don't interleave).

\textbf{Semantic Conflicts (Agent-Mediated)}: CRDTs cannot detect semantic inconsistencies (duplicate declarations, type mismatches, broken references). \textbf{Preliminary measurement}: Manual inspection of 60/600 runs (10\% sample) identified $\sim$5--10\% semantic conflict rate with high task variance (20\% for simple tasks, 80\% for complex tasks); comprehensive measurement across all 600 runs needed for precise rates. \textbf{Resolution}: Evaluator agent identifies conflicts via TypeScript diagnostics; applies automatic fixes or flags for review.

\section{Experimental Methodology}
\label{sec:evaluation}

\subsection{Research Questions}

\textbf{RQ1 (Performance)}: Parallel vs. sequential response time and efficiency.

\textbf{RQ2 (Quality)}: Parallelization impact on code quality, architecture, performance, accessibility.

\textbf{RQ3 (Consistency)}: CRDT-based strong eventual consistency validation.

\subsection{Experimental Design}

\subsubsection{Benchmark Suite and Task Coupling}

Six diverse TypeScript/React tasks spanning complexity spectrum. We characterize tasks by \textbf{coupling}---the degree of interdependency between components, operationalized as the fraction of TODOs whose implementation requires reading or modifying shared state established by other TODOs. High coupling ($>$50\%): Pomodoro Timer (shared timer state, synchronized UI updates), Dashboard (shared data context), Algorithm Visualizer (coordinated animation state). Low coupling ($<$30\%): Tic-Tac-Toe (independent cell logic, isolated game state), Registration Form (independent field validators), Markdown Editor (partially independent formatting functions). Medium coupling (30-50\%): remaining tasks with mixed dependencies. \textbf{Limitation}: Coupling measured post-hoc via manual inspection of generated code dependencies rather than pre-defined metrics; future work should use static analysis (data-flow graphs, shared variable access patterns) for objective coupling measurement.

\subsubsection{Evaluation Procedure}

Each task: 50 runs per mode (sequential, parallel). \textbf{Metrics}: Response time (end-to-end: prompt submission $\rightarrow$ evaluator completion, including CRDT sync); code quality, architecture, performance, accessibility scores (0--20 each). \textbf{Total}: 6 tasks $\times$ 2 modes $\times$ 50 runs = \textbf{600 evaluations}.

\subsubsection{Scoring Methodology}

Claude Sonnet 4.5 evaluates code using rubric: \textbf{Code Quality} (0--20: type safety, patterns, error handling, maintainability), \textbf{Architecture} (0--20: component structure, state management), \textbf{Performance} (0--20: optimization, efficiency), \textbf{Accessibility} (0--20: ARIA, keyboard nav), \textbf{Functionality} (0--20: requirements, correctness).

\subsubsection{Implementation}

Claude Sonnet 4.5 via Amazon Bedrock; 120~s timeout; up to 5 parallel agents; sub-10~ms CRDT sync; SQLite persistence.

\subsection{Statistical Analysis}

\textbf{Outlier Removal}: Per-task-mode IQR method for response time (13.8\% removed overall: 83/600 data points); no removal for scores. \textbf{Hypothesis Testing}: Per-task Wilcoxon signed-rank tests combined via fixed-effects meta-analysis with inverse-variance weighting. Compute Cohen's d\_z (paired effect size), pooled estimates, and $I^{2}$ heterogeneity. Bonferroni correction $\alpha = 0.05/6$ for 6 metrics. \textbf{Power}: n=300/mode provides $>$0.99 power for medium effects (d=0.5).

\subsection{Threats to Validity}

\textbf{Internal}: LLM stochasticity (temp=0 but $\pm$21.57~s variance remains), network variability (13.8\% outliers removed via per-task-mode IQR), LLM-based scoring subjectivity, \textbf{confounded response time measurements}---parallel generates 82--189\% more code for some tasks, conflating code generation volume with coordination overhead; current methodology cannot isolate these factors. \textbf{External}: Task bias (6 UI tasks $<$100 LOC), language limitation (TypeScript/React only), scale (5 agents max), representativeness (no $>$10k LOC codebases). \textbf{Construct}: LLM scoring lacks human baseline; response time includes end-to-end latency not pure compute.

\textbf{Scope and Interpretation}: Our evaluation compares parallel vs. sequential execution within the same observation-driven coordination pattern, isolating benefits of parallelism from CRDT-specific overhead. We do NOT compare CRDTs vs. alternative substrates (OT, replicated logs, file-locking)---our results characterize the \textbf{coordination pattern} (observation-driven with deterministic convergence), not the optimality of CRDTs specifically. Findings about task-dependent scalability likely generalize to other substrates providing observable updates and deterministic convergence, though substrate-specific performance characteristics (metadata overhead, convergence latency, coordination cost) would differ significantly.

\textbf{Mitigation}: Large samples (50 runs/task), diverse complexity, Bonferroni correction, transparent negative results, explicit acknowledgment that results characterize pattern not primitive.

\section{Results and Discussion}
\label{sec:results}

\subsection{Overview}

\textbf{100\% evaluation pipeline completion} across 600 evaluations; zero crashes or data corruption. All runs completed successfully (no evaluation failures, agent crashes, or CRDT synchronization errors). Generated code quality measured separately: TypeScript compilation errors range from 0.59--5.93 per 1000 characters (Table~\ref{tab:objective}); ``pipeline completion'' refers to successful execution of the evaluation workflow, not generated code correctness or compilability. Table~\ref{tab:results} shows meta-analyzed results.

\begin{table*}[t]
\caption{Meta-analyzed results (6 tasks $\times$ 50 runs per mode; per-task Wilcoxon tests with inverse-variance pooling). $I^{2}=0\%$ indicates negligible between-task heterogeneity in \textbf{effect directions} (all tasks show consistent direction within each metric despite magnitude differences), validating pooling for sign determination. Note: $I^{2}$ measures heterogeneity in effect sizes, not raw performance---per-task magnitudes vary dramatically, but $I^{2}=0\%$ means this variance is explained by within-task uncertainty rather than true between-task effect heterogeneity.}
\label{tab:results}
\vskip 0.15in
\centering
\small
\begin{tabular}{lcccccc}
\toprule
\textbf{Metric} & \textbf{Sequential} & \textbf{Parallel} & \textbf{$\Delta$ (s or pts)} & \textbf{$\Delta$ (\%)} & \textbf{p-value} & \textbf{Effect Size (d\_z)} \\
\midrule
Response Time (s) & 60.92~s & 68.90~s & +7.98~s & +13.1\% & 0.022 & 0.13 (negligible) \\
Overall Score & 55.98 & 56.52 & +0.54 & +1.0\% & 0.029 & 0.13 (negligible) \\
Code Quality & 17.13 & 15.81 & -1.32 & -7.7\% & $<$0.001 & -0.71 (medium) \\
Architecture & 13.61 & 13.50 & -0.11 & -0.8\% & 0.158 & -0.08 (negligible) \\
Performance & 11.06 & 13.82 & +2.76 & +25.0\% & $<$0.001 & 1.51 (large) \\
Accessibility & 14.18 & 13.39 & -0.79 & -5.6\% & $<$0.001 & -0.59 (medium) \\
\bottomrule
\end{tabular}
\end{table*}

\subsection{RQ1: Performance---Variable Outcomes}

Parallel 13.1\% slower overall with \textbf{significant task-to-task variation}: ranging from -21.1\% speedup (Tic-Tac-Toe) to +39.4\% slowdown (Algorithm Visualizer). Table~\ref{tab:pertask} shows per-task breakdown.

\begin{table}[t]
\caption{Per-task response time results}
\label{tab:pertask}
\vskip 0.15in
\centering
\small
\begin{tabular}{lccc}
\toprule
\textbf{Task} & \textbf{Seq (s)} & \textbf{Par (s)} & \textbf{$\Delta$ (\%)} \\
\midrule
Tic-Tac-Toe & 45.47 & 35.89 & \textbf{-21.1\%} \\
Registration & 56.13 & 52.15 & \textbf{-7.1\%} \\
Markdown & 69.33 & 64.32 & \textbf{-7.2\%} \\
Pomodoro & 56.27 & 76.47 & \textbf{+35.9\%} \\
Dashboard & 64.43 & 83.19 & \textbf{+29.1\%} \\
Visualizer & 66.39 & 92.57 & \textbf{+39.4\%} \\
\bottomrule
\end{tabular}
\end{table}

\textbf{Analysis}: Observed slowdowns result from confounded factors. \textbf{Normalized time analysis} (per 1000 characters) reveals parallel coordination is faster per character for 5/6 tasks (11--52\% speedup when controlling for code volume). Raw slowdowns are \textbf{artifacts of code generation volume}---parallel produces 82--189\% more code with optimizations/safety checks.

\subsection{RQ2: Code Quality---Mixed Results}

\textbf{Mixed results}: Parallel excels at performance (+25\%, p$<$0.001, d\_z=1.51) but degrades code quality (-7.7\%, p$<$0.001, d\_z=-0.71) and accessibility (-5.6\%, p$<$0.001, d\_z=-0.59). Architecture unchanged (p=0.158).

\textbf{Objective Metrics}: Static analysis (TypeScript errors, code length) on all 600 samples shows task-dependent outcomes (Table~\ref{tab:objective}).

\begin{table*}[t]
\caption{Per-task objective metrics (Mann-Whitney U tests; ***p$<$0.001, **p$<$0.01, *p$<$0.05). Highly task-dependent: 5/6 tasks show -46\% to -87\% error rate reductions; Markdown Editor worsens (+24\%). Code length: -12\% to +189\%. Critical Correlation: All tasks with highest code inflation (Pomodoro, Dashboard, Visualizer: +82--189\% code) exhibit largest slowdowns (+29--39\%), confounding attribution to coordination overhead alone.}
\label{tab:objective}
\vskip 0.15in
\centering
\small
\begin{tabular}{lcclc}
\toprule
\textbf{Task} & \multicolumn{2}{c}{\textbf{TS Errors/1k chars}} & \multicolumn{2}{c}{\textbf{Code Length (chars)}} \\
& \textbf{Seq} & \textbf{Par ($\Delta$\%)} & \textbf{Seq} & \textbf{Par ($\Delta$\%)} \\
\midrule
Tic-Tac-Toe & 4.60 & 0.59 \textbf{(-87\%)***} & 12,930 & 11,509 (-11\%)*** \\
Registration & 4.24 & 2.27 \textbf{(-47\%)*} & 17,196 & 18,885 (+10\%)*** \\
Pomodoro & 4.09 & 1.95 \textbf{(-52\%)*} & 14,952 & 27,195 \textbf{(+82\%***)} \\
Markdown & 3.26 & 4.04 \textbf{(+24\%)} & 18,285 & 16,028 (-12\%)*** \\
Dashboard & 5.93 & 1.91 \textbf{(-68\%)*} & 19,194 & 37,922 \textbf{(+98\%***)} \\
Visualizer & 3.78 & 0.92 \textbf{(-76\%)*} & 18,389 & 53,068 \textbf{(+189\%***)} \\
\bottomrule
\end{tabular}
\end{table*}

\textbf{Analysis}: Normalized error rates show parallel benefits decomposable tasks but hurts coupled tasks (Markdown Editor outlier). Code length inflates dramatically for complex tasks (+82--189\%: Pomodoro +81.9\%, Dashboard +97.6\%, Visualizer +188.6\%). \textbf{Critical Correlation}: All tasks with highest code inflation exhibit largest slowdowns (+29--39\%), confounding attribution to coordination overhead alone. \textbf{Correlation with LLM scores}: Fewer TS errors (syntactic) but lower Code Quality scores (semantic)---parallel optimizes for compilability over maintainability. Code inflation co-occurs with +25\% Performance score (likely optimization logic). \textbf{Limitations}: Captures syntactic correctness only; missing semantic/deep quality. Response time measurements conflate code generation volume with coordination overhead.

\subsection{RQ3: Consistency---Zero Data Corruption}

Zero data corruption. \textbf{CRDT guarantees validated}: Strong eventual consistency (all edits converged identically), deterministic conflict resolution, no character-level data loss. Convergence $<$200~ms (5-agent stress test).

\section{Discussion}
\label{sec:discussion}

\subsection{When Observation-Driven Coordination Works}

\textbf{Normalized time analysis reveals parallel coordination is faster per character for 5/6 tasks} (11--52\% speedup when controlling for code volume), demonstrating coordination effectiveness. Raw response times mislead: apparent slowdowns result from code generation volume (parallel produces 82--189\% more code with optimizations/safety checks), not coordination inefficiency. Tasks with independent components benefit most (Visualizer: 51.8\% faster per character).

\textbf{When It Faces Challenges}: Highly coupled tasks (Markdown Editor) show true coordination overhead (+5.8\% slower per character). Code volume inflation from independent TODO optimization creates trade-off: more robust/optimized code (correlates with +25\% Performance scores) but longer generation time. Accessibility-critical apps suffer from isolated agent decisions; SLA-bound systems cannot tolerate high variance ($\pm$21.57~s).

\textbf{vs. Existing Approaches}: ChatDev/MetaGPT use waterfall/pipeline workflows without concurrent execution (predictable but sequential); CodeCRDT enables parallel coordination with per-character efficiency gains despite volume increases. Strong eventual consistency eliminates character-level merge failures---0\% character-level conflicts requiring manual resolution, though edits interleave and converge deterministically. Note: This differs from Git branch merging (structural conflicts requiring manual intervention) vs. concurrent editing (character-level conflicts). Semantic conflicts (preliminary estimate 5--10\% from 60-run inspection) require reconciliation regardless of primitive choice.

\subsection{Scalability and Limitations}

\textbf{Scalability Projections}: Three overhead sources: CRDT metadata O(N$\times$O) (manageable to N=50), observation processing O(N$\times$U) (dominant at N $\approx$ 25-30), context invalidation O(N$\times$k) (k invalidations per agent causing thrashing at N$>$10 for interdependent tasks). \textbf{Amdahl's Law}: Tasks showing speedups peak at N=3 (2.05$\times$), degrade to break-even at N $\approx$ 20; tasks with slowdowns degrade immediately. \textbf{Optimal}: 3--5 agents for suitable tasks; sequential for others. Table~\ref{tab:scalability} shows projected scalability inflection points.

\begin{table*}[t]
\caption{Scalability inflection points (speculative projections beyond N=5)}
\label{tab:scalability}
\vskip 0.15in
\centering
\small
\begin{tabular}{lcccc}
\toprule
\textbf{Metric} & \textbf{1--3} & \textbf{5--10} & \textbf{20--30} & \textbf{50+} \\
\midrule
CRDT Metadata & Low & Low & Moderate & High \\
Observation O(N$\times$U) & Low & Moderate & \textbf{Dominant} & Prohibitive \\
Context Invalidation & Rare & Occasional & Frequent & Constant \\
Speedup (best-case) & 1.5--2.0$\times$ & 1.2--1.4$\times$ & 1.0--1.2$\times$ & $<$1.0$\times$ \\
Speedup (worst-case) & 0.9--1.0$\times$ & 0.7--0.9$\times$ & $<$0.5$\times$ & $<$0.3$\times$ \\
\bottomrule
\end{tabular}
\end{table*}

\textbf{Limitations}: (1) \textbf{Pattern vs. primitive}: Evaluation characterizes observation-driven coordination pattern, not comparative performance of CRDTs vs. OT/consensus-based alternatives---primitive-specific trade-offs (latency, metadata overhead, convergence properties) remain unexplored; (2) \textbf{Code generation volume confound} (addressed via normalized time analysis): Raw response time conflates code volume with coordination overhead; normalized analysis reveals 5/6 tasks show per-character speedup, isolating true coordination costs, though LLM latency variability ($\pm$21.57~s) remains; (3) LLM scoring only (no human validation); (4) Semantic conflicts measured via preliminary 60/600 inspection only (comprehensive measurement needed); (5) 5 agents max (scalability projections beyond N=5 speculative); (6) TypeScript/React only (generalization to other languages/domains unexplored); (7) High variance unsuitable for SLA-bound systems.

\textbf{Future Work}: Understand why parallel generates more code (independent optimization, safety hedging), primitive comparison (CRDTs vs. OT vs. consensus), semantic conflict detection (LSP integration, AST-level coordination), scalability sweeps (N=1--20), multi-language, runtime testing (functional correctness, actual performance), formal verification, variance reduction.

\section{Conclusion}
\label{sec:conclusion}

CodeCRDT applies observation-driven coordination---a decades-old pattern from distributed systems---to concurrent multi-agent LLM code generation. Building on prior work in Linda tuplespaces, blackboard architectures, and stigmergy, we formalize this pattern for stochastic LLM agents with provable safety guarantees and empirically characterize when parallel coordination succeeds vs. fails based on task structure.

\textbf{Contributions}: (1) Formalization and implementation of observation-driven coordination for stochastic LLM agents with provable safety (TODO-claim protocol ensuring at-most-one agent per task under strong eventual consistency); (2) 600-trial evaluation with normalized time analysis revealing parallel is faster per character for 5/6 tasks (11--52\% speedup when controlling for code volume), demonstrating that apparent slowdowns result from code generation volume rather than coordination inefficiency; (3) Demonstration that LLM agents exhibit distinct failure modes: semantic conflicts despite character-level convergence, code volume inflation from local optimization, quality-performance trade-offs; (4) Open-source implementation, deployment heuristics, and methodological insights on measuring parallel LLM coordination when code generation volume confounds response time.

\textbf{Key Findings}: Raw response times show task-dependent outcomes (21.1\% speedup to 39.4\% slowdown), but normalized time analysis reveals parallel is faster per character for 5/6 tasks (11--52\% speedup when controlling for code volume). Apparent slowdowns are artifacts of code generation volume---parallel produces 82--189\% more code with optimizations/safety checks, requiring more inference time but generated more efficiently. Only highly coupled tasks (Markdown Editor: +5.8\% slower per character) show true coordination overhead dominance. Quality trade-offs: +25\% performance optimization correlating with code volume inflation, but -7.7\% code quality, -5.6\% accessibility from isolated agent decisions. Results validate observation-driven coordination effectiveness while revealing emergent behavior: parallel agents optimize locally, producing more robust but verbose code.

\textbf{Impact}: Observation-driven coordination enables decentralized AI collaboration with formal consistency guarantees, with variable effectiveness depending on task characteristics. The pattern generalizes beyond CRDTs to any substrate providing observable updates and deterministic convergence. Applications beyond code generation: collaborative reasoning, multi-modal generation, edge AI. As agent-based systems proliferate, observation-driven coordination offers principled foundation with empirically characterized trade-offs.

\clearpage

\bibliography{references-mlsys}
\bibliographystyle{mlsys2025}

\appendix

\section{Implementation Details}
\label{app:implementation}

\subsection{CRDT Configuration}

Hocuspocus WebSocket relay with 15~s sync timeout, exponential backoff (1~s--30~s), max 5 reconnection attempts.

\subsection{Agent Prompts (Abbreviated)}

Outliner generates TypeScript/React skeletons with TODO markers; Implementation agents fill TODOs using cursor tool; Evaluator scores quality/architecture/performance/accessibility (0--20 each).

\subsection{Statistical Analysis}

Per-task Wilcoxon signed-rank tests with Mann-Whitney U for objective metrics; Cohen's d\_z effect sizes; Bonferroni correction $\alpha$ = 0.05/6.

\subsection{Effect Size Explanation}

Table~\ref{tab:results} reports Cohen's d\_z (paired differences) rather than Cohen's d (independent samples). The switch from unpaired (d$\approx$0.5) to paired analysis (d\_z=1.51 for Performance) increases effect sizes by removing between-task variance. Per-task pairing properly accounts for experimental design where same tasks run in both modes.

\subsection{TODO-Claim Protocol: Formal Specification}

\textbf{State}: Shared Y.Map with entries \texttt{TODO\_k = \{description, status $\in$ \{pending, claimed, done\}, assignedTo $\in$ AgentID $\cup$ \{null\}, logicalClock\}}

\textbf{Claim Operation} (agent A):

\begin{algorithmic}[1]
\STATE \textbf{claim}(A, k):
\STATE \quad \textbf{pre}: TODO\_k.status = pending $\land$ TODO\_k.assignedTo = null
\STATE \quad 1. write(TODO\_k.assignedTo $\leftarrow$ A, logicalClock $\leftarrow$ A.counter++)
\STATE \quad 2. wait(SYNC\_DELAY)  // 50~ms for CRDT convergence
\STATE \quad 3. v $\leftarrow$ read(TODO\_k.assignedTo)
\STATE \quad 4. \textbf{return} (v == A)  // claim succeeded iff we won
\end{algorithmic}

\textbf{Safety Theorem}: At any point after convergence, $\forall$TODO\_k: $|\{A : A.claimSucceeded(k)\}| \leq 1$

\textbf{Proof Sketch}:
\begin{enumerate}
\item Yjs Y.Map implements LWW register per key (assignedTo is a key)
\item Concurrent writes resolve via (logicalClock, clientID) lexicographic total order
\item All replicas converge to same winner (SEC guarantee)
\item Verify-after-wait ensures agents observe post-convergence state
\item Therefore, at most one agent observes \texttt{assignedTo == self} post-convergence
\end{enumerate}

\textbf{Liveness}: Eventually, all pending TODOs are claimed (assuming $\geq$1 live agent, finite retry, no infinite failures). Crashed agents' stale claims reclaimed via 120~s timeout + status reset.

\textbf{Idempotency}: Re-claiming already-claimed TODO is safe (verify step fails, agent moves to next).

\textbf{ABA Prevention}: Logical clocks monotonically increase per agent; clientID tie-breaking is deterministic.

\section{Statistical Tables}
\label{app:stats}

\subsection{Normalized Response Time Analysis}

To isolate coordination overhead from code generation volume, we compute response time normalized by code length (seconds per 1000 characters):

\begin{table}[t]
\caption{Response time normalized by code length (s/1000 chars)}
\label{tab:normalized}
\vskip 0.15in
\centering
\small
\begin{tabular}{lccc}
\toprule
\textbf{Task} & \textbf{Seq} & \textbf{Par} & \textbf{$\Delta$ (\%)} \\
\midrule
Tic-Tac-Toe & 3.52 & 3.12 & \textbf{-11.4\%} \\
Registration & 3.26 & 2.76 & \textbf{-15.3\%} \\
Markdown & 3.79 & 4.01 & \textbf{+5.8\%} \\
Pomodoro & 3.76 & 2.81 & \textbf{-25.3\%} \\
Dashboard & 3.36 & 2.19 & \textbf{-34.8\%} \\
Visualizer & 3.61 & 1.74 & \textbf{-51.8\%} \\
\bottomrule
\end{tabular}
\end{table}

\textbf{Critical Finding}: When controlling for code length, parallel coordination is faster per character for 5/6 tasks (11--52\% speedup). Raw slowdowns result from code generation volume (parallel produces 82--189\% more code), not coordination inefficiency.

\textbf{Implications}: (1) Observation overhead O(N$\times$U) is \textbf{not} the dominant factor for 5/6 tasks; (2) Code generation volume inflation drives raw response time; (3) Parallel agents generate code more efficiently (faster per character) despite producing more code; (4) Only highly coupled tasks (Markdown Editor) show true coordination overhead dominance.

\subsection{Outlier Analysis}

83/600 data points removed (13.8\%) via per-task-mode IQR method. Breakdown by task: Algorithm Visualizer (7 outliers: 3 parallel, 4 sequential), Dashboard (5: 3 parallel, 2 sequential), Markdown Editor (37: 15 parallel, 22 sequential), Pomodoro Timer (14: 10 parallel, 4 sequential), Registration Page (13: 7 parallel, 6 sequential), Tic-Tac-Toe (7: 3 parallel, 4 sequential). Symmetric overall across modes (42 parallel, 41 sequential). Causes: network timeouts, Amazon Bedrock throttling, system load. Robustness checks recommended: no-removal baseline, winsorization, quantile regression. \textbf{Per-Task Detailed Results}: Available in supplementary materials with mean, std, 95\% CI per task and mode.

\end{document}